\documentclass[12pt]{article}
\usepackage[top=1in, left=1in, right=1in, bottom=1in]{geometry}	
\usepackage[parfill]{parskip}	% Activate to begin paragraphs with an empty line rather than an indent
\usepackage{graphicx}
\usepackage{amssymb}
\usepackage{epstopdf}
\usepackage{bbm}
\usepackage{amssymb}
\usepackage{amsmath}
\usepackage{amsfonts}

\usepackage{lscape}
\usepackage{rotating}
\usepackage{setspace}
\usepackage{threeparttable}
\usepackage{booktabs}
\usepackage{subcaption}
\usepackage[compact]{titlesec}
\usepackage{lmodern}
\fontfamily{lmtt}\selectfont
\usepackage[T1]{fontenc}

\usepackage{natbib}
\bibpunct{(}{)}{;}{a}{}{,}
\usepackage{hyperref}
\usepackage{titling}
\newcommand{\subtitle}[1]{%
  \posttitle{%
    \par\end{center}
    \begin{center}\large#1\end{center}
    \vskip0.5em}%
}
\usepackage{amsthm}

\usepackage{color}
\begin{document}
\pagestyle{plain}

\newcommand{\blind}{0}

\newcommand{\tit}{\Large Notes on Causation, Comparison, and Regression}

\if0\blind

{\title{\tit
\thanks{We thank Isaiah Andrews, Kosuke Imai, Ronald Kessler, Russell Localio, Michael Sobel, Joseph Newhouse, Bijan Niknam, and Paul Rosenbaum for helpful comments and conversations. For graphics, we thank Xavier Alema\~ny and Esmeralda Aceituno.
This work was supported by the Alfred P. Sloan Foundation (G-2020-13946) and the Patient Centered Outcomes Research Initiative (PCORI, ME-2022C1-25648).}\vspace*{.3in}}
\author{Ambarish Chattopadhyay
\thanks{Stanford Data Science, Stanford University, 450 Jane Stanford Way Wallenberg, Stanford, CA 94305; email: \url{hsirabma@stanford.edu}.}
\and Jos\'{e} R. Zubizarreta\thanks{Departments of Health Care Policy, Biostatistics, and Statistics, Harvard University, 180 Longwood Avenue, Office 307-D, Boston, MA 02115; email: \url{zubizarreta@hcp.med.harvard.edu}.}
}

\date{}

%\date{First draft: March 27, 2011 
%\\ 
%\medskip
%This draft: \today}

\maketitle
}\fi

\if1\blind
\title{\bf \tit}
\maketitle
\fi

\begin{abstract}
Comparison and contrast are the basic means to unveil causation and learn which treatments work. To build good comparison groups, randomized experimentation is key, yet often infeasible. In such non-experimental settings, we illustrate and discuss diagnostics to assess how well the common linear regression approach to causal inference approximates desirable features of randomized experiments, such as covariate balance, study representativeness, interpolated estimation, and unweighted analyses. We also discuss alternative regression modeling, weighting, and matching approaches and argue they should be given strong consideration in empirical work.
\end{abstract}

\vspace*{.3in}

\begin{center}
\noindent Keywords: 
%\small 
{Causal Inference; Randomized Experiments; Observational Studies; Regression Modeling; Weighting Adjustments; Multivariate Matching}
%\normalsize
\end{center}
\clearpage
\doublespacing

\singlespacing
\pagebreak
\tableofcontents
\pagebreak
\doublespacing

%%%%%%%
%%%%%%%
%%%%%%%
%%%%%%%
%%%%%%%
%%%%%%%
\section{Unweighted comparisons work well in randomized experiments}
\label{sec1}

%%%%%%%
%%%%%%%
\subsection{Learning which treatments work}

Comparison and contrast are the basic means to learn the effects of causes --- which treatments work --- in a complex world of which we only know so much. 
Especially in fields like medicine and the social sciences, where the distinct heterogeneity of each individual makes them unique, making these comparisons is crucial and requires meticulous attention.
But to render these contrasts useful and `see' the effect of treatment, it is crucial that subjects exposed and not exposed to the treatment are similar. This is essential to isolate the effect from observable and unobservable confounding factors. 

%%%%%%%
%%%%%%%
\subsection{Randomized comparisons}

The randomized experiment stands out as the ideal method for building such comparable groups. Randomization tends to produce groups with similar pre-treatment observed and unobserved characteristics that differ only on the receipt of treatment, and possibly, if the treatment is effective, on their subsequent outcomes. Randomization also provides a factual basis to quantify whether differences in the outcomes are due to system or chance and to conduct statistical tests for the effect of treatment \citep{fisher1935design}. To illustrate, consider the following simulated example inspired by the famous RAND Health Insurance Experiment \citep{joseph1993free}.\footnote{The data and code of this example are available upon request.}

%%%%%%%
%%%%%%%
\subsection{Running example}

The goal of the investigation is to learn the causal effect of a new health insurance plan (the \emph{treatment} variable) on health expenditures (the \emph{outcome} variable). The purpose of this new plan is to protect patients against high out-of-pocket medical costs while reducing health spending, so we anticipate that this plan will lead to a decrease in average health expenditure. The basic parameter we wish to learn is the average reduction in health expenditure caused by the insurance plan among those who enroll; that is, the average effect of treatment on the treated subjects, henceforth denoted ATT. We focus on the ATT as opposed to the average effect of treatment on all subjects (ATE) for illustrative purposes, but our analyses can be readily extended to encompass the ATE.

To this end, suppose that a randomized experiment is conducted on a random sample of 200 individuals from a population. Half of these individuals are randomly assigned to enroll in the plan --- the treatment group --- and the other 100 are similarly assigned to a control group. For each individual, there is information on two pre-treatment characteristics, or covariates; namely, household per-capita income and number of hospital visits in the previous year. 
Table \ref{tab:example} presents a sample of data from this experiment for ten typical individuals in each group. 

%\hspace{-2cm}
\begin{table}[h!]
\caption{20 typical individuals in the randomized experiment. The baseline covariates are Income and Visits. The treatment variable is Insurance (1 if insured, 0 otherwise). The outcome variable is Expenditure.}
\hspace{.5cm}
    %\centering
    %\scalebox{1}{
    \begin{center}
    \begin{tabular}{rrrr}
    \hline
        Income & Visits & Insurance & Expenditure\\
        \hline
37451 &   1 &   1 & 6250 \\ 
48509 &   6 &   1 & 8461 \\ 
31202 &   1 &   1 & 5234 \\ 
52205 &  10 &   1 & 9356 \\ 
49854 &   4 &   1 & 8522 \\ 
24977 &   0 &   1 & 4134 \\ 
51224 &   2 &   1 & 8613 \\ 
26432 &   5 &   1 & 4734 \\ 
52012 &   5 &   1 & 8940 \\ 
50009 &   0 &   1 & 8254 \\ 
37749 &   1 &   0 & 7658 \\ 
41520 &   6 &   0 & 8903 \\ 
32736 &   5 &   0 & 7053 \\ 
52025 &   3 &   0 & 10696 \\ 
45111 &   4 &   0 & 9427 \\ 
25673 &   0 &   0 & 5144 \\ 
52462 &   2 &   0 & 10700 \\ 
27404 &   5 &   0 & 5964 \\ 
45305 &   5 &   0 & 9560 \\ 
50938 &   0 &   0 & 10182 \\ 
  \hline
    \end{tabular}
   % }
   \end{center}
    \label{tab:example}
\end{table}

\begin{figure}[h!]
\caption{Sketch of a randomized experiment. Covariate values are displayed over the shoulder of each individual. The diagnostic dashboard shows that randomization has produced balance in the covariates between the groups and in relation to the target population without weighting. As a result, the effective sample size equals the nominal sample size, and the weights are constant and equal to one. Here, the target population is the treated population because the estimand is the average treatment effect among the treated.}
\begin{center}
\includegraphics[scale=.75]{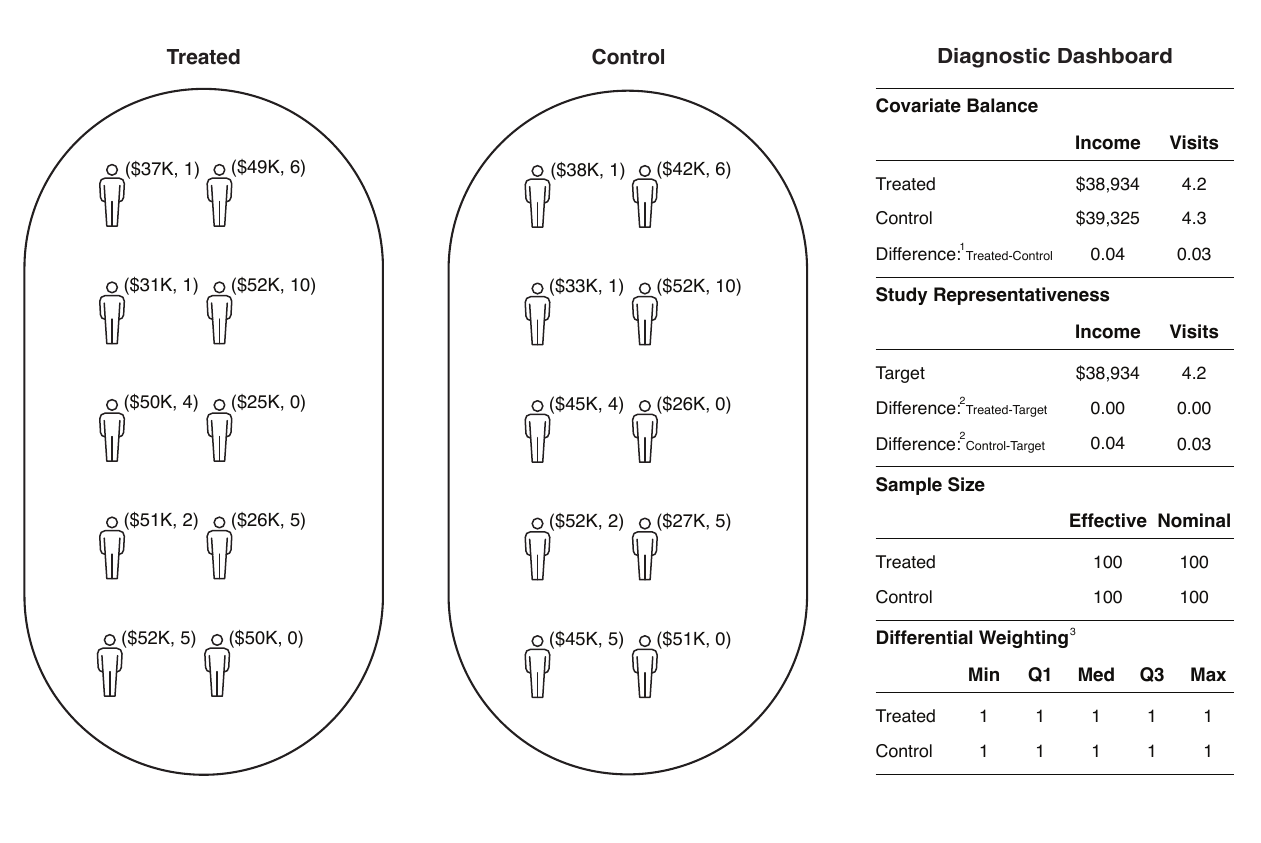}
\end{center}
\vspace{-.5cm}
\subcaption*{\footnotesize{$^1$: Absolute standardized mean difference. $^2$: Target absolute standardized mean difference. $^3$: The weights are scaled to sum up the size of each treatment group. Min: minimum. Q1, Q3: first and third quartiles. Med: median. Max: maximum.}}
\label{fig_rand}
\end{figure}

Figure \ref{fig_rand} portrays the randomized experiment for these 20 individuals. 
In the figure, the covariate values for each individual are shown over their shoulder. For example, the treated individual in the top-left corner has a household per-capita income of \$37K and 1 hospital visit in the previous year. The \emph{diagnostic dashboard} on the rightmost panel tabulates first, under `covariate balance,' the means of these two covariates in the sample of 100 treated subjects, 100 control subjects, and their absolute standardized mean difference \citep{rosenbaum1985constructing}.\footnote{The absolute standardized mean difference of a covariate $x$ between the treatment and the control group is defined as $\frac{|\bar{x}_t - \bar{x}_c|}{\sqrt{(s^2_t + s^2_c)/2}}$, where $(\bar{x}_t,s_t)$ and $(\bar{x}_c,s_c)$ are the mean and standard deviations of $x$ in the treatment and control group, respectively. Smaller values of this measure indicate better mean balance on $x$.}
This diagnostic shows that, save for chance imbalances, the groups bear close resemblance to each other in terms of the covariate means.
Second, under `study representativeness,' it shows the corresponding means in the target (here, treated) population as well as the absolute standardized mean differences between each group and the target. Third, under `sample size,' it presents the effective sample size (i.e., the number of subjects effectively used for estimation) and the nominal sample size (i.e., the original number of subjects) in each group.\footnote{More formally, for a sample of $n$ subjects with weights $w_1,w_2,...,w_n$, the effective sample size is defined as $\frac{(|w_1| + |w_2| + ... + |w_n|)^2}{w^2_1 + w^2_2 + ... + w^2_n}$. If the weights are uniform, the effective sample size equals the nominal sample size $n$.} 
Finally, under `differential weighting,' it displays the minimum, maximum, and quartiles (25th, 50th, and 75th percentiles) of the distribution of the weights (in this case, constant).  
This diagnostic quantifies the deviation from the ideal scenario of uniform (i.e., constant) weights which aid interpretability and minimize the variance of general weighting estimators.

%%%%%%%
%%%%%%%
\subsection{Weighted contrasts}

After randomization, because the groups tend to have similar observed and unobserved characteristics, we can estimate the average effect of treatment by simply taking the mean difference in outcome values $y$ between the treated and the control groups,

$$\overbrace{\frac{y_{t,1} + y_{t,2} + ... + y_{t,10}}{10}}^{\text{Treated group}} - \overbrace{\frac{y_{c,1} + y_{c,2} + ... + y_{c,10}}{10}}^{\text{Control group}}.$$
\hspace{.25cm}

In this expression, each observation contributes equally to the total effect estimate, so we can say that the treated and control groups are unweighted or uniformly weighted (because each observation has a constant weight equal to one) and equivalently write

$$\overbrace{\frac{1 \cdot y_{t,1} + 1 \cdot y_{t,2} + ... + 1 \cdot y_{t,10}}{10}}^{\text{Treated group}} - \overbrace{\frac{1 \cdot y_{c,1} + 1 \cdot y_{c,2} + ... + 1 \cdot y_{c,10}}{10}}^{\text{Control group}}.$$
\hspace{.25cm}

For our simulated example, the average difference in the outcome (health expenditures) between the treatment and control group is roughly -\$1.5K, suggesting the insurance plan reduces health expenditure by that amount.

In randomized experiments, without weights the treated and control groups are balanced in expectation, so in principle, we can estimate the average effect of treatment by contrasting the simple averages in the treated and control groups. This observation is important as we move forward to more complex settings and methods.

%%%%%%%
%%%%%%%
\subsection{Observational studies}

 The power of randomized experiments stems from its treatment assignment mechanism --- which is controlled by the investigator --- and by its resulting data structure, which is transparent and simple: by design, we can clearly see how randomization works, whether the randomly formed groups are comparable or not, and we can talk about it, fostering collaboration and the elaboration of scientific theories. However, due to ethical or practical constraints, we cannot always carry out experiments. In these situations, we must fall back on observational studies, where by means of a process unbeknownst to the investigator, subjects are selected into treatment or control, typically in an unbalanced way. 

\begin{figure}[h!]
\caption{Sketch of an observational study before adjustments. Covariate values are displayed alongside each individual's shoulder. The diagnostic dashboard shows covariate imbalances among groups and in comparison to the target population. At this stage, no differential weighting is applied, and the effective sample size equals to the nominal sample size.}
\begin{center}
\includegraphics[scale=.75]{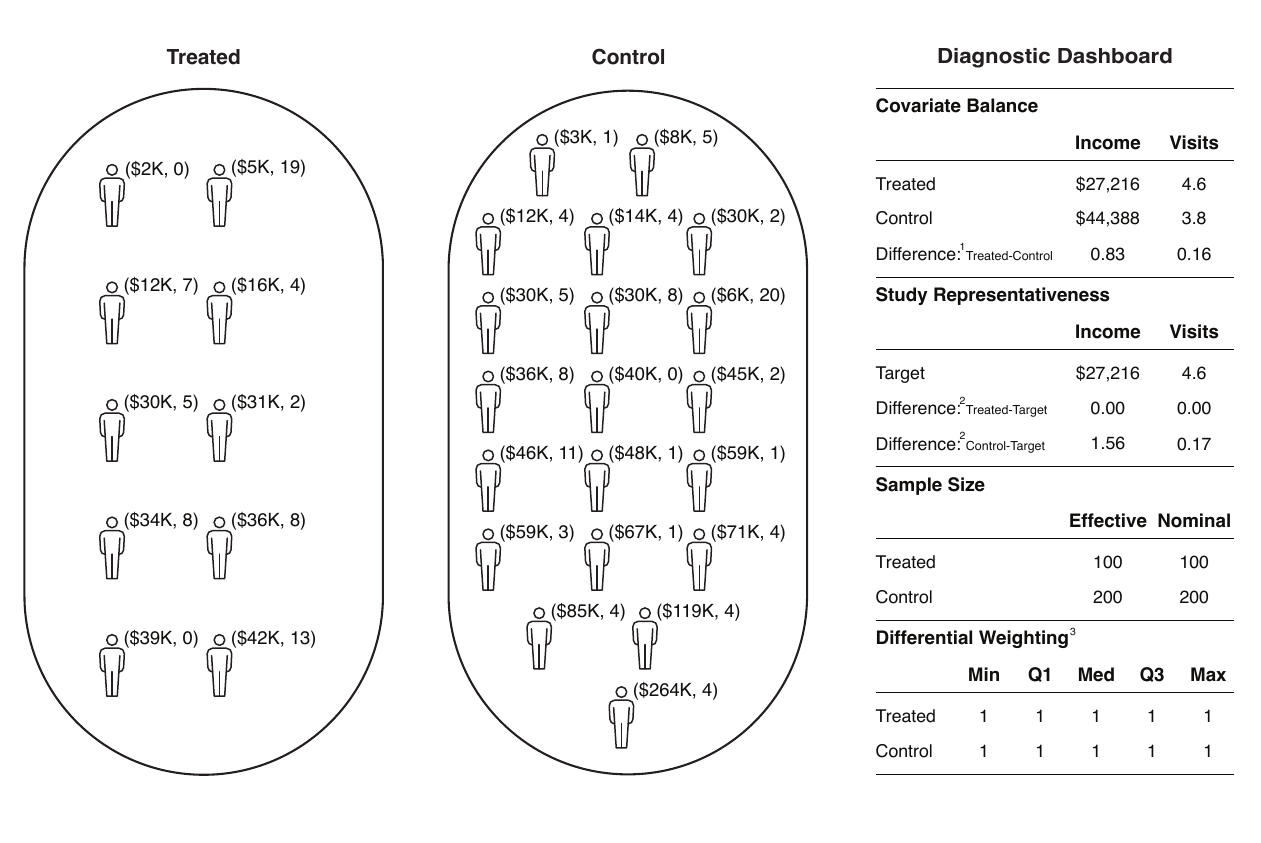}
\end{center}
\vspace{-.5cm}
\subcaption*{\footnotesize{$^1$: Absolute standardized mean difference. $^2$: Target absolute standardized mean difference. $^3$: The weights are scaled to sum up the size of each treatment group. Min: minimum. Q1, Q3: first and third quartiles. Med: median. Max: maximum.}}
\label{fig_obs_before}
\end{figure}

Following our running example, we now turn our attention to a parallel observational study. 
This is a simulated study of 300 subjects randomly drawn from a population distinct from the previous randomized experiment. 
From this population, a process unknown to the investigator has led some subjects to enroll in the new health insurance plan, while others have not. 
As before, the goal is to learn the average treatment effect on the treated or ATT. In the sample, there are 100 treated subjects and 200 controls. Figure \ref{fig_obs_before} summarizes this for 10 and 20 typical subjects from the treated and control groups, respectively. The average difference in the outcome variable, health expenditures, between the treatment and control group is roughly -\$5K, suggesting that the insurance plan reduces the health expenditure by that amount. But how trustworthy is this conclusion?

%\vspace{-.5cm}

 As seen in the diagnostic dashboard of Figure \ref{fig_obs_before}, the treatment groups differ in their observed characteristics, and likely in the unobserved ones. For instance, the average income in the treated group is \$27.2K, which is markedly different from the \$44.4K average income among the controls, likely implying that the observed differences in health expenditure are driven by this pre-treatment imbalance. Therefore, to estimate the average treatment effect on the treated, some form of adjustments must be made to make the groups comparable, at least in terms of the observed characteristics. To this end, the randomized experiment is the conceptual benchmark that guides most methods of adjustment in an observational study (\citealt{dorn1953philosophy}; \citealt{cochran1965planning}; \citealt{rosenbaum2010design1}; \citealt{imbens2015causal}; \citealt{hernan2020causal}).

%%%%%%%
%%%%%%%
%%%%%%%
\section{Regression models implicitly use unequal weights for \\ estimating treatment effects}

%%%%%%%
%%%%%%%
\subsection{Implied adjustments}

In observational studies we seek to approximate or emulate as closely as possible randomized experiments in order to estimate the effect of treatments \citep{rubin2007design}. However, observational studies often rely on regression models to adjust for covariate imbalance and estimate such effects. In these analyses, how well does regression mimic key features of a randomized experiment? Specifically, how does linear regression recover (i) covariate balance, (ii) study representativeness, (iii) interpolated estimation, and (iv) unweighted samples that facilitate transparent analyses?

%%%%%%%
%%%%%%%
\subsection{Approximating experiments}

In this question, (i) covariate balance refers to making the observed characteristics between the treatment and control groups similar. The influence of likely imbalances in unobserved characteristics must be assessed with a sensitivity analysis (e.g., \citealt{cornfield1959smoking}; \citealt{rosenbaum1987sensitivity, rosenbaum2005sensitivity}; \citealt{vanderweele2017sensitivity}; \citealt{zhao2019sensitivity}). (ii) Study representativeness is the requirement that the characteristics of each group be similar to (and hence, representative of) a target population. (iii) Interpolated estimation or sample-boundedness means estimating effects that are an interpolation of the observed data rather than an extrapolation beyond the range or support of the actual data \citep{robins2007comment}. (iv) Finally, an unweighted or self-weighted sample refers to a simple data structure where the observations have homogeneous weights and which does not require further adjustments that can decrease the efficiency and transparency of the study. A recent paper provides answers to this question by deriving and analyzing the implied weights of linear regression \citep{chattopadhyay2023implied}.

%%%%%%%
%%%%%%%
%%%%%%%
\section{Regression adjustments in nonexperimental comparisons}

It is common to fit regression models to nonexperimental data, including linear regression, logistic regression, and proportional hazards regression. Consider the following linear regression model where the outcome of interest $y$ is regressed on observed covariates or characteristics (e.g., age, gender, and risk factors) and the treatment assignment indicator,\footnote{The treatment assignment indicator of a subject equals one if the subject receives treatment and equals zero otherwise.}
$$\text{Outcome variable} = \alpha + \beta \cdot \text{Observed covariates}+ \tau \cdot \text{Treatment indicator} + \epsilon.$$
In the model, $\varepsilon$ is a random error term with mean zero, distributed mean-independently of the covariates and treatment, and $\alpha$, $\beta$ and $\tau$ are the unknown regression coefficients. Under the assumption that the observed covariates encompass all relevant confounders, the objective again is to estimate the parameter $\tau$ for the ATT. 
This parameter is typically estimated by ordinary least squares (OLS) from a random sample of observations from a large (possibly infinite) population, where it is postulated that uncertainty stems from the randomness of the sampling process.

Arguably, this is the traditional regression approach to causal inference. By standard regression theory, if the assumed model is correct, then the traditional OLS estimator is the most efficient, linear unbiased estimator for the ATT. But where is the experiment? Does the regression approach approximate the above features of randomized experiments? And what happens if the model is used when it is incorrect?

%%%%%%%
%%%%%%%
%%%%%%%
\section{A regression adjustment is implicitly a weighted comparison}

To answer these questions, we must first ask a crucial intermediate question. How does the above regression approach implicitly weight the individual observations in the data at hand? Building on important work by \citet{imbens2015matching} (see also \citealt{abadie2015comparative}, \citealt{gelman2018high}), \citet{chattopadhyay2023implied} show that the traditional regression estimator of $\tau$ is equivalent to the following weighted contrast
\begin{equation}
\overbrace{\frac{w_{t,1} \cdot y_{t,1} + w_{t,2} \cdot y_{t,2} + ... + w_{t,10} \cdot y_{t,10}}{10}}^{\text{Treated group}} - \overbrace{\frac{w_{c,1} \cdot y_{c,1} + w_{c,2} \cdot y_{c,2} + ... + w_{c,10} \cdot y_{c,10}}{10}}^{\text{Control group}}
\label{eq:weightedcomparison}
\end{equation}
where the weights $w$ in each group sum to the corresponding group size and can be expressed in closed form; i.e., without approximations.\footnote{See \citet{angrist1999empirical} and \citet{aronow2016does} for related work.} If, in the control group, $w_{c,1} = 2$, then the first control counts as if she were two controls, correcting for the fact that people like this control are underrepresented in the control group and overrepresented in the treated group because of the absence of randomization. In other words, linear regression implicitly creates and contrasts weighted samples of treated and control observations, where the contribution or weight of each observation to the aggregate average treatment effect estimate can be computed and scrutinized. By analyzing these weights, we can transparently assess how linear regression attempts to mimic features of randomized experiments, as we discuss next. In Figure \ref{fig_obs_uri}, we illustrate the structure of the weighted groups after regression in our running example.\footnote{Here, we regress the outcome (expenditure) linearly on the two baseline covariates: income and visits.} The implied regression weights are computed using the \texttt{R} package \texttt{lmw}; see \citet{chattopadhyay2023lmw} for details. Here, the size of individuals is proportional to their corresponding weights, with negative weights turning them upside down.

%%%%%%%
%%%%%%%
%%%%%%%
\section{The implied regression weights create a limited form of covariate balance}

%%%%%%%
%%%%%%%
\subsection{Exact covariate balance.}

\begin{figure}[h!]
\caption{Sketch of an observational study after regression adjustments. Covariate values are displayed alongside the shoulder of each individual. The size of each person is scaled according to their contribution or weight in the calculation of the average treatment effect. Individuals with a negative weight are depicted upside down.}
\begin{center}
\includegraphics[scale=.75]{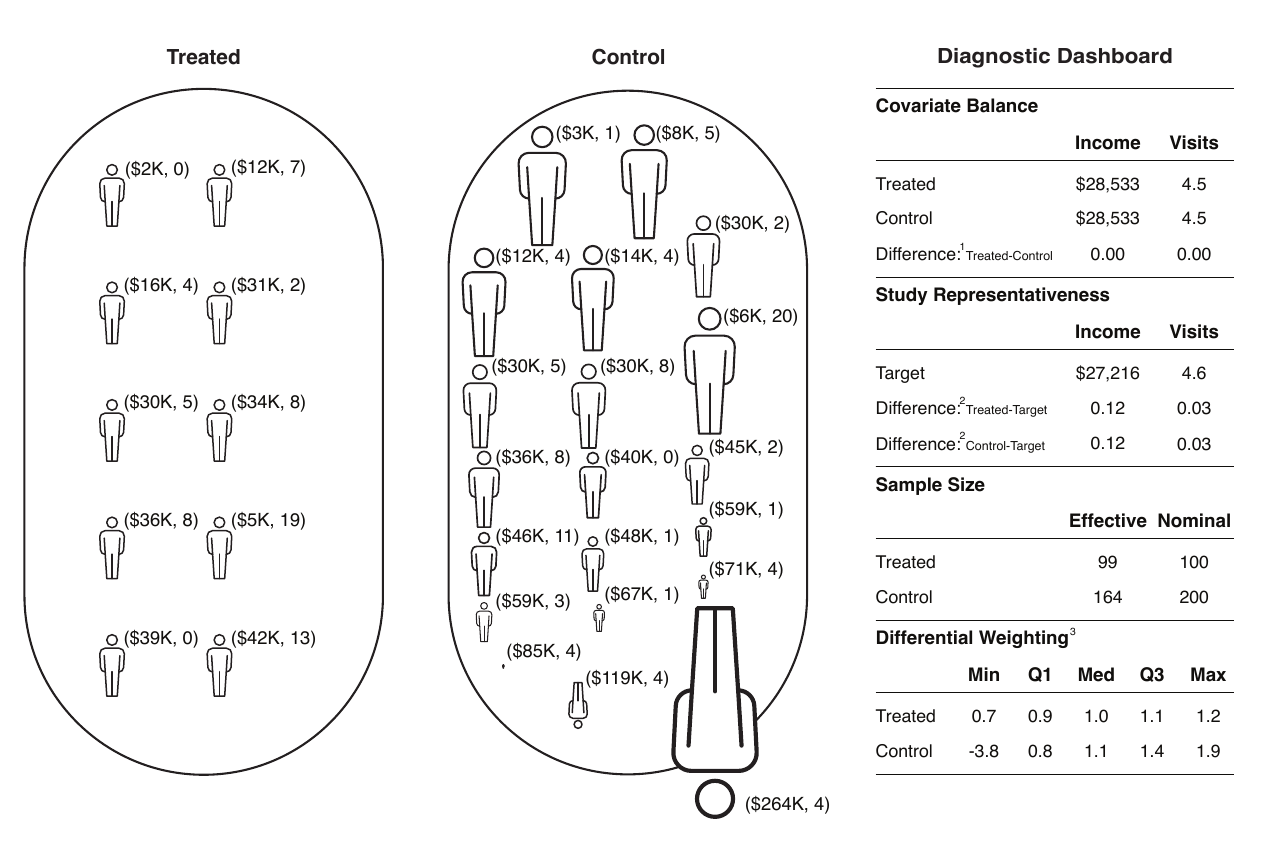}
\end{center}
\vspace{-.3cm}
\subcaption*{\footnotesize{$^1$: Absolute standardized mean difference. $^2$: Target absolute standardized mean difference. $^3$: The weights are scaled to sum up the size of each treatment group. Min: minimum. Q1, Q3: first and third quartiles. Med: median. Max: maximum.}}
\label{fig_obs_uri}
\end{figure}

Does linear regression balance covariates between the treated and control groups? As it turns out, there is a specific form of covariate balance concealed within the linear regression model. The implied weights of linear regression exactly balance the means of the covariates included in the model, such that the means of those observed characteristics are identical between the weighted treated and weighted control groups. See the diagnostic dashboard in Figure \ref{fig_obs_uri}: across the two groups, the mean pre-treatment income and hospital visits are the same after regression adjustments. Furthermore, if one includes transformations of the covariates in the regression model, such as squares or interactions between them, then the means of these transformations will also be exactly balanced. However, there is no guarantee of balance for covariate transformations that are not included in the model.\footnote{Unless the true propensity score model is inverse linear on the covariates included in the model (\citealt{robins2007comment}; \citealt{chattopadhyay2023implied}).}

%%%%%%%
%%%%%%%
\subsection{A hidden population.}

Does linear regression create treated and control samples that are representative of the target population? While the above linear regression model exactly balances the means of the covariates in the treated and control samples, it may not balance them at the target population. See the diagnostic dashboard in Figure \ref{fig_obs_uri}. After regression, the average income in both groups is \$28.5K, whereas the average income in the target (i.e., treated) population is \$27.2K. Similarly, after regression, the average number of hospital visits in both groups is 4.5, which differs from 4.6, the average in the target. 

In general, linear regression may not balance the samples relative to the natural populations characterized by the overall subjects, the treated subjects, or the control subjects. Implicitly, it balances them elsewhere, at a different covariate profile (see \citealt{chattopadhyay2023implied}, for an exact formula). In this example, the profile is (\$28.5K, 4.5), which does not correspond to any of these natural populations. This profile may in fact characterize a population that does not exist. In other words, here linear regression ends up estimating the average treatment effect on a hypothetical population of individuals whose average income is \$28.5K and who has 4.6 hospital visits on average. 

Taking a step back, the above linear regression model assumes that the treatment effect is constant everywhere. If this assumption is correct, then it does not really matter which population regression ends up targeting: the average treatment effect is the same as that in any other population. But if this assumption is incorrect, then effect estimates will be biased, especially if the implied weights take large negative values, as we explain next. 

%%%%%%%
%%%%%%%
\subsection{Estimation beyond the support of the data and negative weights.}

Does linear regression produce interpolated or sample-bounded estimates \citep{robins2007comment} of the treatment effects? To answer this question, we first note that the implied weights of the regression model can take negative values. In Figure \ref{fig_obs_uri}, subjects with negative weights are upside down. The most notable of them is a control subject with an income of \$264K and 4 hospital visits, who is assigned a very large negative weight. 

From a practical standpoint, negative weights are hard to interpret: while a person with a weight value of 1/2 in the sample means that it has half its weight in the population, and a weight of zero removes the person from the population, what is the meaning of a person with a weight value of -1? From an analytic perspective, these negative weights translate into forming effect estimates that may lie outside the support or range of the observed data, and thus that are not sample-bounded. In other words, one can get estimates that are very distant from any of the measurements in the actual data. As a result, one may end up obtaining rather silly estimates of the treatment effect. In our example, the effect estimate under regression is \$147, suggesting that health insurance increases health expenditure. 
If the assumed regression model is correct, then such extrapolation is not detrimental in large samples, because the regression estimator is consistent for the ATT.
However, if the model is incorrect, then this problem may arise even in large samples.  

%%%%%%%
%%%%%%%
\subsection{Differential weighting of minimum variance.}

Does linear regression produce an evenly weighted or self-weighted sample? In general, unless the covariate means are exactly balanced between the treated and control groups before adjustment, regression ensures such mean balance by enforcing differential weighting of individuals. In our example, differential weighting is evident in the control group and is also present in the treatment group. As shown in the diagnostic dashboard in Figure \ref{fig_obs_uri}, this differential weighting results in effective sample sizes that are smaller than the nominal sample sizes.

Further, an additional fundamental property is that the implied regression weights have minimum variance. Thus, although the sample individuals are differentially weighted, the weights are `as stable as possible' subject to their mean balancing property.\footnote{In other words, the OLS estimator for $\tau$ can be equivalently obtained solving an optimization problem that minimizes the variance of the weights that exactly balance the means of the covariates between the treatment and control groups, with weights that add up to one and that can take negative values. See \citet{chattopadhyay2023implied} for a general expression of this optimization problem that also encompasses weighted least squares (WLS), augmented inverse probability weighting (AIPW), and ridge regression estimation, in connection to the stable balancing weights (SBW; \citealt{zubizarreta2015stable}).} Hence, the OLS estimator of $\tau$ is the best (minimum variance) linear unbiased estimator if the assumed regression model is correct. The issue, again, is that the true model that generates the data is unknown. If it were known, we wouldn't need the data. If the assumed model is incorrect, then this stability property of the regression weights offers limited value, because in that case, regression may end up balancing the wrong functions, targeting the wrong population, and extrapolating beyond the range of the data. 

%%%%%%%
%%%%%%%
%%%%%%%
\section{Alternative regression modeling, weighting, and matching approaches}

%%%%%%%
%%%%%%%
\subsection{Multi regression.}

An alternative regression approach extends the previous model by including interactions between the observed covariates and the treatment assignment indicator as follows,\footnote{In this case, the covariates need to be centered, i.e., have mean zero. This can be achieved by subtracting from each covariate its sample mean before fitting the model. Centering allows us to represent the average treatment effect as a single coefficient $\tau$, as opposed to a combination of $\tau$ and $\gamma$.}
\begin{align*}
    \text{Outcome variable} &= \alpha + \beta \cdot \text{Observed covariates} + \tau \cdot \text{Treatment indicator} \\
    & \hspace{0.4cm} + \gamma \cdot \text{Observed covariates} \cdot \text{Treatment indicator} + \epsilon .
\end{align*}
This approach can be used to estimate the average treatment effect in the overall population, which corresponds to the parameter $\tau$. See \citet{peters1941method} and \citet{belson1956technique} for early discussions of this method. By including the treatment-covariates interactions, this regression model specification now allows for effect modification or heterogeneity. As discussed by \citet{chattopadhyay2023implied}, this approach is equivalent to fitting separate linear regression models of the outcome on the covariates in the treated and control groups, and then taking the mean difference in the predicted outcomes under treatment and control, based on the two fitted models.

For estimating the ATT the multi regression approach proceeds as follows. First fit the regression model
$$\text{Outcome variable} = \alpha + \beta \cdot \text{Observed covariates} + \epsilon$$
in the \emph{contro}l group. Then, for each unit in the treatment group, predict its outcome in the absence of treatment, by using the previously fitted regression model. Finally, over all the treated units, compute the mean of their observed outcome minus the mean of the predicted outcomes.

\begin{figure}[h!]
\caption{Sketch of an observational study after multi or interacted regression adjustments. The covariate values for each individual are shown over their shoulder. The size of each person is scaled according to their contribution or weight in the calculation of the average treatment effect. Individuals with a negative weight are depicted upside down.}
\begin{center}
\includegraphics[scale=.75]{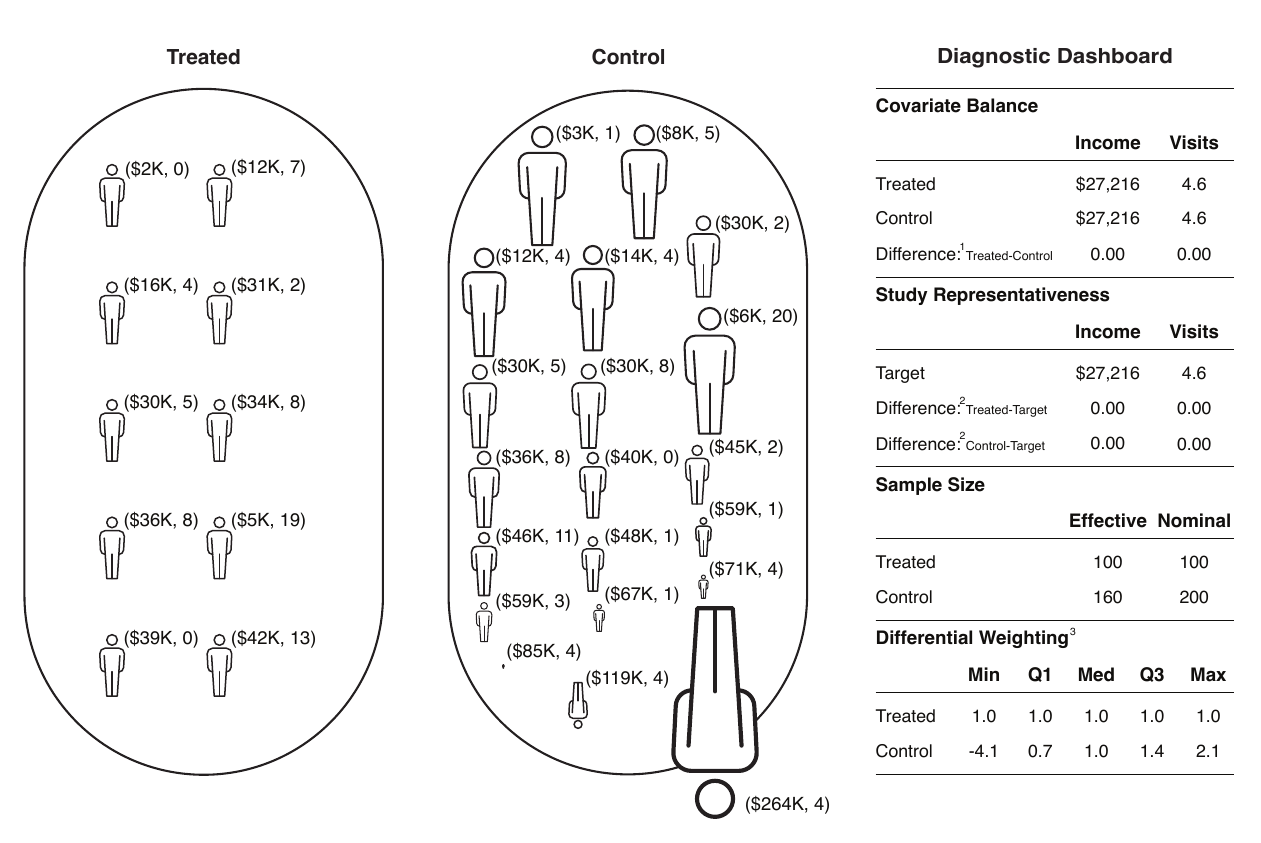}
\end{center}
\vspace{-.3cm}
\subcaption*{\footnotesize{$^1$: Absolute standardized mean difference. $^2$: Target absolute standardized mean difference. $^3$: The weights are scaled to sum up the size of each treatment group. Min: minimum. Q1, Q3: first and third quartiles. Med: median. Max: maximum.}}
\label{fig_obs_mri}
\end{figure}

To what extent does this approach approximate an experiment? \citet{chattopadhyay2023implied} show that, under this model specification, the estimator of the ATT can also be written as a weighted contrast, 
$$\overbrace{\frac{1 \cdot y_{t,1} + 1 \cdot y_{t,2} + ... + 1 \cdot y_{t,10}}{10}}^{\text{Treated group}} - \overbrace{\frac{w_{c,1} \cdot y_{c,1} + w_{c,2} \cdot y_{c,2} + ... + w_{c,10} \cdot y_{c,10}}{10}}^{\text{Control group}}.$$
with implied weights $w$ that satisfy the following properties. (i) Like the previous approach, the means of the observed covariates are exactly balanced between the treatment and control groups. (ii) The covariate means of each group are centered at the treated population (the target population for the ATT), addressing the representativeness problem of the previous approach. (iii) An issue that persists is that the weights can take negative values, leaving room for biases due to extrapolation from an incorrectly specified model. (iv) As before, regression induces differential weighting on the control subjects such that the weights have the minimum variance among those that exactly balance the means of the covariates at the target population. See Figure \ref{fig_obs_mri} for an illustration of these properties.

Properties (i) and (ii) are manifest because the covariate profiles of both groups are (\$27.2K, 4.6), the same as that of the target. Also, the differential weighting in property (iv) is evident in the control group, while by construction, subjects are weighted uniformly in the treatment group. Finally, similar to the previous regression approach, property (iii) is present in the control group, with the same control subject having an income of \$264K and 4 hospital visits getting a very large negative weight. The effect estimate turns out to be \$309 which, like the previous approach, contradicts established substantive knowledge.  

Through the lens of the above properties, this multi regression approach comes closer to approximating the ideal randomized experiment than the previous approach. Thus, unless there is a high degree of certainty that treatment effects are homogeneous, the multi regression approach may be preferred. However, a potential drawback that persists is the bias that may arise from extrapolating a wrong model, coupled with a loss in interpretative simplicity due to departures from an unweighted sample.

Arguably, the performance of both regression methods discussed can be improved by conducting careful exploratory analyses of the data before adjustment and estimation. For instance, in our example, diagnostics may suggest that the control unit with profile (\$264K, 4) is an outlier in terms of income and may be discarded before adjustment. However, such checks can be overlooked in routine use of regression in practice, and moreover, with multiple covariates, these checks may not be straightforward. Ideally, one would use methods that have some of these checks built-in, and hence, that force the investigator to carefully inspect the data. Along these lines, we now discuss weighting and matching methods for adjustments in observational studies.

\begin{figure}[h!]
\caption{Sketch of an observational study after weighting adjustments (SBW). The covariate values for each individual are shown over their shoulder. The size of each individual is proportional to his/her contribution or weight in the average treatment effect estimate. Individuals with a weight of zero are depicted in grey.}
\begin{center}
\includegraphics[scale=.75]{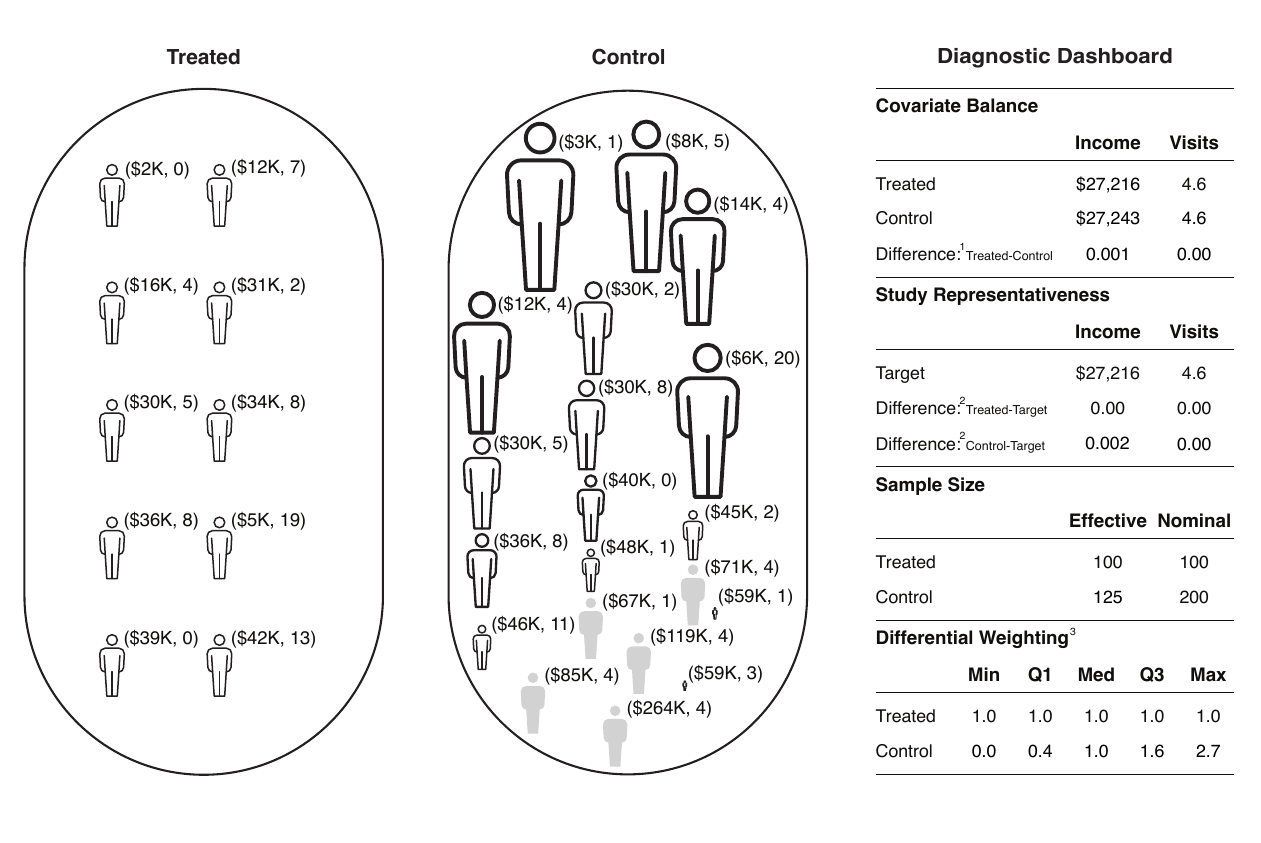}
\end{center}
\vspace{-.5cm}
\subcaption*{\footnotesize{$^1$: Absolute standardized mean difference. $^2$: Target absolute standardized mean difference. $^3$: The weights are scaled to sum up the size of each treatment group. Min: minimum. Q1, Q3: first and third quartiles. Med: median. Max: maximum.}}
\label{fig_obs_sbw}
\end{figure}

%%%%%%%
%%%%%%%
\subsection{Weighting.}

In the two aforementioned regression methods, we started with an outcome model and then considered its implied weights to understand how these methods attempt to mimic a randomized experiment. Alternatively, acknowledging that the true outcome model is unknown, we can instead consider directly weighting the treatment and control groups without explicitly specifying an outcome model \citep{rosenbaum1987model}. The most commonly used weighting method is based on modeling the probability of treatment assignment, or propensity score, using the observed covariates \citep{rosenbaum1983central}.\footnote{More formally, for estimating the ATT, a subject with covariate $x$ receives a weight of 1 if treated and a weight $\frac{\hat{e}(x)}{1-\hat{e}(x)}$ if not treated, where $\hat{e}(x)$ is the value of the estimated propensity score for the subject.} If the model for the propensity score is correct, then in large samples we can unbiasedly estimate the average treatment effect by simply weighting the observations by the inverse of their propensity score estimates. 

Does the above weighting procedure create a sample that looks as if it came from a randomized experiment? For a correctly specified propensity score model, (i) the weights balance, in large samples, the entire joint distribution of the observed covariates between the two groups. (ii) They also balance each group relative to the target (i.e., treated) population. With an incorrect model, however, large imbalances between the groups and relative to the target may exist for some covariates or their transformations. (iii) The propensity score is, by definition, non-negative and hence, for any reasonable propensity score model (correct or incorrect) the resulting effect estimates are sample-bounded. (iv) The resulting weights can be extreme and highly variable across individuals, especially when there is a lack of overlap in covariate distributions between the treatment and control groups. 

Overall, weighting is a powerful and flexible method to adjust for covariates. However, in view of the experimental ideal, care must be taken to adequately balance the covariates toward the right target population, with weights that produce stable estimators. Recent weighting methods that emphasize these features are \citet{hainmueller2012balancing}, \citet{imai2014covariate}, and \citet{zubizarreta2015stable}, among others. See \citet{austin2015moving} and \citet{chattopadhyay2020balancing} for reviews.

In our example, we implement the stable balancing weights (SBW) method of \citet{zubizarreta2015stable}. Figure \ref{fig_obs_sbw} shows the structure of the resulting weighted sample. Overall, the covariates are reasonably well-balanced relative to the target population, although not exactly (see the balance diagnostic for income). By construction, for the ATT, this method puts differential weights only on the control units, but with weights that are non-negative and less extreme than the regression methods. Notice that it puts zero weight on the outlying control subject with profile (\$264K, 4). With this method, the estimated treatment effect amounts to -\$967 in the example. This estimate aligns with existing substantive knowledge, and is closer to the estimate from the randomized experiment. We note that these results need not be identical, because the data for the randomized experiment and the observational study were drawn from different populations. When treatment effects vary across populations (i.e., there is effect modification), these results do not necessarily have to coincide.

%%%%%%%
%%%%%%%
\subsection{Matching.}

Matching is a common approach that aims to find the randomized experiment that is `hidden inside' \citep{hansen2004full} the observational study (\citealt{stuart2010matching}; \citealt{imbens2015matching}; \citealt{rosenbaum2020modern}). In its simplest form, matching forms pairs of similar treated and control subjects, such that the matched groups are balanced on aggregate. After matching, the structure of the data is similar to that of a randomized experiment: (i) covariates --- the observed ones --- are transparently balanced, while the unobserved ones, like with any other method for adjustment, must be dealt with a sensitivity analysis. (ii) Representativeness is overt and easily verifiable, as the target population can simply be compared to the matched sample in balance tables. (iii) By achieving covariate balance, effect estimates mostly are an interpolation grounded in the actual data and not an extrapolation of a potentially incorrect model. (iv) Each observation in the matched sample has the same weight, so the simplest test statistics and graphical displays can be used to analyze the outcomes. Finally, conducting a sensitivity analysis after matching is straightforward using Rosenbaum bounds (\citealt{rosenbaum1987sensitivity}; \citealt{rosenbaum2002observational}, chapter 4). Popular matching methods include \citet{rosenbaum1989optimal}, \citet{iacus2011causal}, and \citet{diamond2013genetic}. Recent optimal matching methods that capitalize on modern optimization using network flows include \citet{hansen2006optimal}, \citet{pimentel2015large}, \citet{yu2020matching}, and \citet{yu2022graded}. For recent optimization matching methods that leverage modern optimization using generic integer programming, see \citet{zubizarreta2012using}, \citet{zubizarreta2014matching}, and \citet{cohn2022profile}.

\begin{figure}[h!]
\caption{Sketch of an observational study after pair matching. The covariate values for each individual are shown over their shoulder. The size of each individual is proportional to his/her contribution or weight in the average treatment effect estimate. Individuals with a weight of zero are depicted in grey. Matched individuals are connected.}
\begin{center}
\includegraphics[scale=.75]{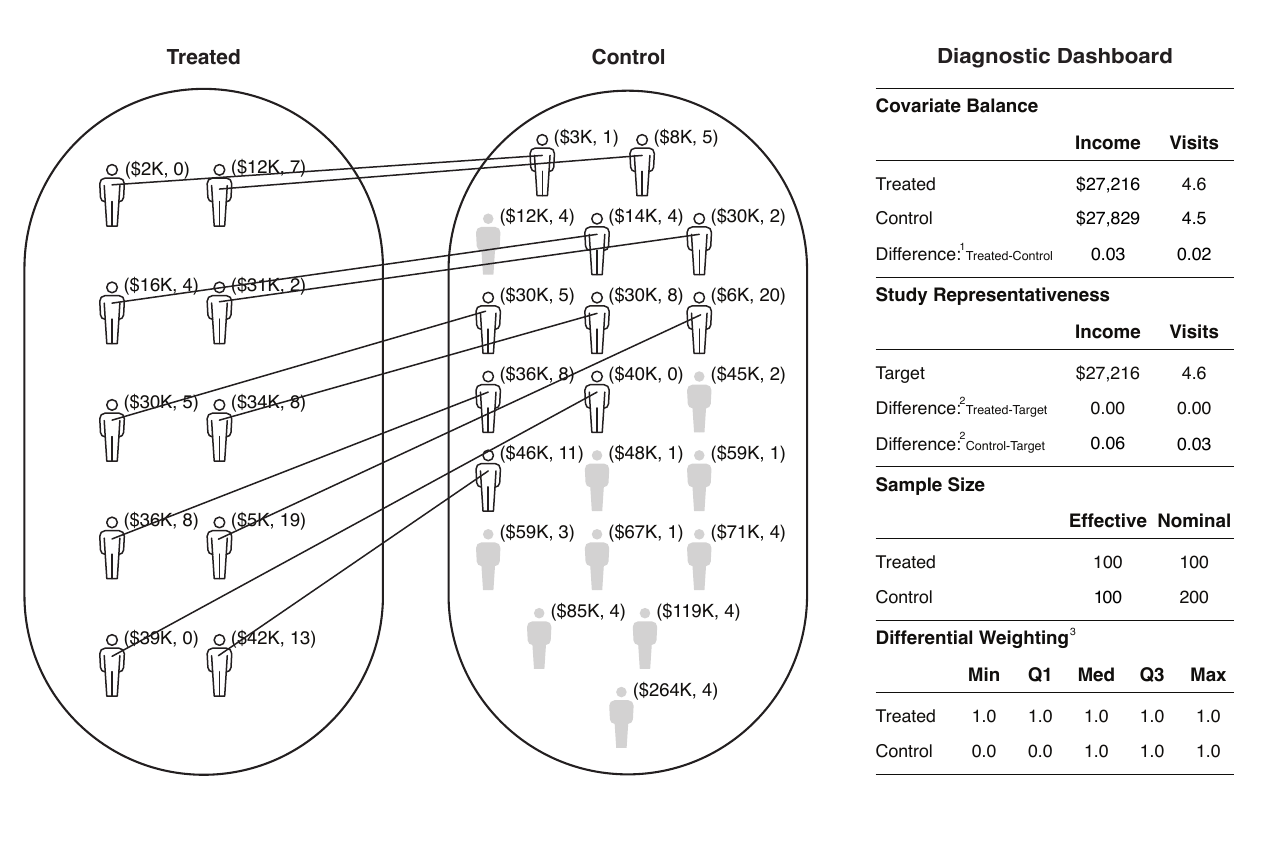}
\end{center}
\vspace{-.5cm}
\subcaption*{\footnotesize{$^1$: Absolute standardized mean difference. $^2$: Target absolute standardized mean difference. $^3$: The weights are scaled to sum up the size of each treatment group. Min: minimum. Q1, Q3: first and third quartiles. Med: median. Max: maximum.}}
\label{fig_obs_pairmatch}
\end{figure}

\begin{figure}[h!]
\caption{Sketch of an observational study after profile matching. The covariate values for each individual are shown over their shoulder. The size of each individual is proportional to his/her contribution or weight in the average treatment effect estimate. Individuals with a weight of zero are depicted in grey.}
\begin{center}
\includegraphics[scale=.75]{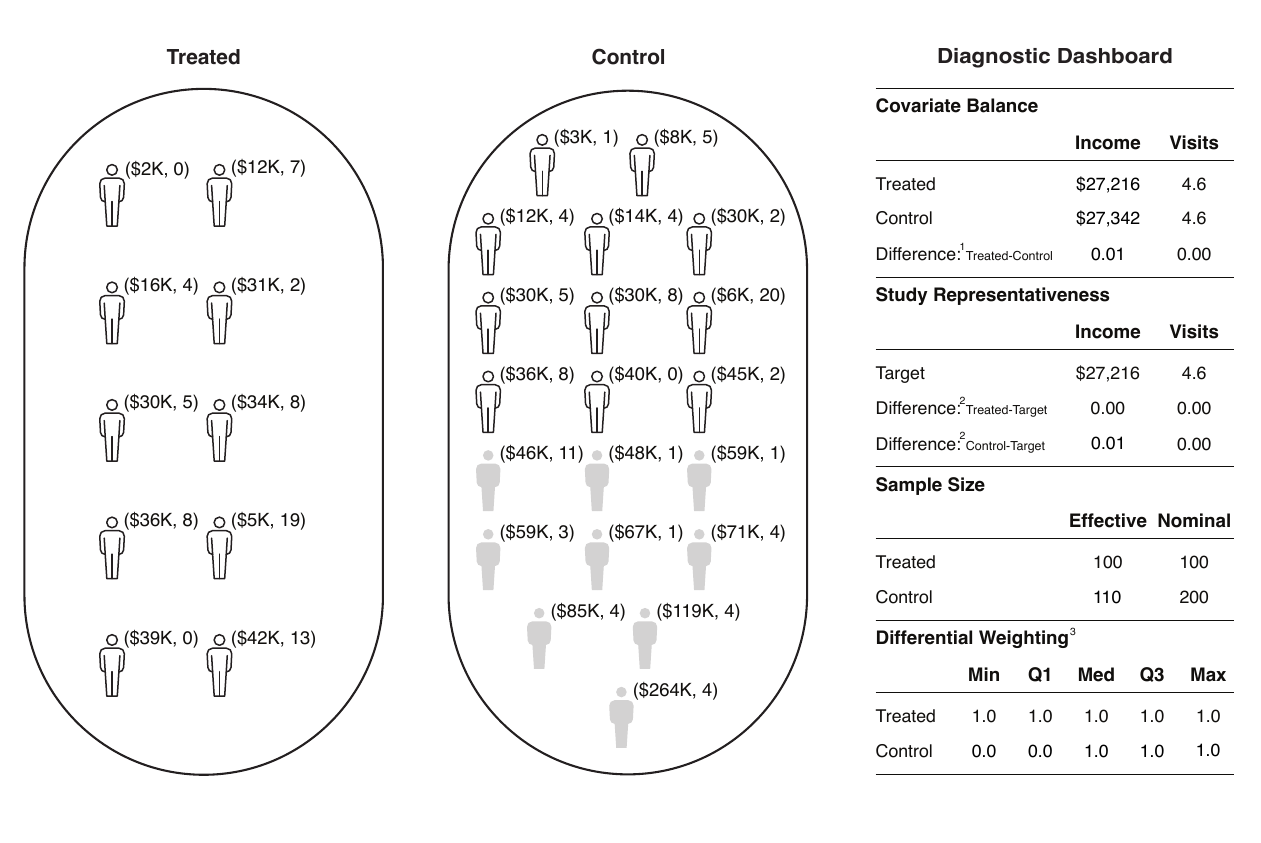}
\end{center}
\vspace{-.5cm}
\subcaption*{\footnotesize{$^1$: Absolute standardized mean difference. $^2$: Target absolute standardized mean difference. $^3$: The weights are scaled to sum up the size of each treatment group. Min: minimum. Q1, Q3: first and third quartiles. Med: median. Max: maximum.}}
\label{fig_obs_profilematch}
\end{figure}

In our example, we implement two matching methods, optimal pair matching as proposed by \citet{rosenbaum1989optimal} and profile matching by \citet{cohn2022profile}. The corresponding structures of the matched samples are depicted in Figures \ref{fig_obs_pairmatch} and \ref{fig_obs_profilematch}. Overall, both methods produce well-balanced and uniformly weighted samples relative to the target population. Finally, the effect estimates under pair matching and profile matching are -\$1124 and -\$999, respectively, close to the estimate provided by SBW. 

%%%%%%%
%%%%%%%
%%%%%%%
%\pagebreak
\section{It's through contrast that we see}

Comparison and contrast are the basic means for understanding whether a treatment works; and randomization, the ideal mechanism for building similar groups of subjects where the effect of treatment is isolated. In the absence of randomization, covariate imbalance adjustments are needed to ensure the groups are comparable, and three core methods that attempt to achieve this are regression, weighting, and matching.

These methods can be enhanced using non-parametric machine learning and combined in doubly robust estimators (Robins et al., \citeyear{robins1994estimation}). For example, as discussed in \citeauthor{chattopadhyay2023implied} (2023, Section 6), some of these doubly robust estimators can be represented as a simpler weighted contrast, as in Equation (\ref{eq:weightedcomparison}). This representation, in turn, allows us to inspect the implied unit-level weights and conduct diagnostic analyses to evaluate covariate balance, study representativeness, effective sample size, differential weighting, and influence.

As discussed, the extent to which matching, regression, and weighting methods approximate the experimental ideal varies and is not always clear-cut. In observational studies, the essence of regression lies in correctly specifying the true model, in which case the approach is optimal. However, true models are elusive, and although some estimated models are helpful, for causal inference, it is often more practical to concentrate first on building good contrasts. Among others, this means approximating an experiment as closely as possible, with guarantees of robustness, interpretability, and ease to assess generalizability to target populations and sensitivity to hidden biases.

Along these dimensions, the traditional regression method is dominated by the multi regression approach, and matching and weighting approaches should be given strong consideration. We hope that increased use of implied weights diagnostics will improve the practice of causal inference. Upon further development, the diagnostics illustrated in this paper may contribute to a better understanding of advanced statistical and machine learning techniques for causal inference. The ultimate goal? More transparent and reliable scientific findings.

%%%%%%%
%%%%%%%
%%%%%%%
%\pagebreak
%\bibliography{ref}
\bibliography{bib2022}
\bibliographystyle{asa}
  
%%%%%%%
%%%%%%%
%%%%%%%

\end{document}